\documentstyle[preprint,aps]{revtex}

\begin{document}
\draft
\tightenlines

\title{The Three-Nucleon System Near the N-d Threshold}

\author{ A.~Kievsky$^1$, S.~Rosati$^{1,2}$, M.Viviani$^1$,
         C.~R.~Brune$^3$, H.~J.~Karwowski$^3$, E.~J.~Ludwig$^3$, 
         and M.~H.~Wood$^3$}
\address{ $^1$Istituto Nazionale di Fisica Nucleare, Piazza Torricelli 2,
          56100 Pisa, Italy }
\address{ $^2$Dipartimento di Fisica, Universita' di Pisa, Piazza Torricelli 2,
          56100 Pisa, Italy }
\address{ $^3$Department of Physics and Astronomy, University of North 
          Carolina at Chapel Hill, Chapel Hill, North Carolina 27599-3255
          and Triangle Universities Nuclear Laboratory, Durham, 
          North Carolina 27708 }

\date{\today}

\maketitle

\abstract{ The three-nucleon system is studied at energies a few
hundred keV above the N-d threshold.
Measurements of the tensor analyzing powers $T_{20}$
and $T_{21}$ for p-d elastic scattering at
$E_{c.m.}=432$ keV are presented
together with the corresponding theoretical predictions. The
calculations are extended to very low energies since they are useful for
extracting the p-d scattering lengths from the experimental data.
The interaction considered here is the Argonne V18 potential plus the
Urbana three-nucleon potential. The calculation of the
asymptotic $D$- to $S$-state ratio for $^3$H and $^3$He, for which recent
experimental results are available, is also presented.}

\narrowtext
\noindent{key words: N-d scattering, polarization observables,
          effective range expansion, asymptotic constants}\\

\newpage
Study of the three-nucleon system at low energies provides a
stringent test of our understanding of the nuclear dynamics and
the nucleon-nucleon (NN) interaction.
In Ref.~\cite{KRTV96}, detailed
comparisons have been performed between experimental data and the
corresponding theoretical predictions for N-d scattering for
$0.6\le E_{c.m.} \le 2.0$~MeV. Good
agreement was observed for the cross section and the tensor analyzing
powers $T_{20}$, $T_{21}$, and $T_{22}$, but significant differences were
found in the vector analyzing power $A_y$ for N-d scattering
and the analyzing power $iT_{11}$ for p-d scattering.
It is also important to experimentally test the
theoretical calculations at lower energies.
One motivation is to extract the p-d scattering lengths,
for which past experimental results~\cite{huttel}
have persistently disagreed with theoretical estimates~\cite{FRIAR89}.
A further advantage is that at low energies the contributions from
higher partial waves are reduced, so that the remaining important
partial waves can be better investigated.
Another motivation for these comparisons is to test
the theoretical p-d scattering wavefunctions which have recently been used
to calculate the astrophysical $S$-factor and analyzing powers for the
${}^2{\rm H}(p,\gamma){}^3{\rm He}$ reaction at very
low energies~\cite{cattura}. 

Below $E_{c.m.}=0.6$~MeV, the cross section is
mainly governed by $S$-wave scattering.
As a result the polarization observables,
which are influenced by other partial waves,
are small and difficult to determine experimentally.
The present paper reports measurements of
$T_{20}$ and $T_{21}$  for p-d elastic
scattering  at  $E_{c.m.}=432$~keV.
These data are compared to calculations
utilizing the Pair-Correlated Hyperspherical Harmonic
(PHH) basis~\cite{KVR93} to construct the scattering wave function.
The corresponding reactance matrix (${\cal R}$ matrix)
is obtained by means of the Kohn variational principle,
as described in Ref.~\cite{KVR94}.
The calculations have been done using the AV18~\cite{AV18}
potential plus the three-nucleon interaction (TNI)
of Urbana (UR)~\cite{Urbana}.
They have been extended to lower energies in order
to investigate the behavior of the ${\cal R}$-matrix
elements near zero energy, and to evaluate
the p-d scattering lengths $^2a_{pd}$ and $^4a_{pd}$.

The asymptotic $S$- and $D$-state constants for which recent accurate results
are available~\cite{Beata,Zeid,Knutson} have also been calculated.
These quantities involve the interaction between three nucleons with
total angular momentum $J^\pi={1\over 2}^+$, which is
a very important partial wave in low-energy N-d scattering.
The properties of the three-nucleon bound states, such as binding energy and
asymptotic constants, and low-energy elastic scattering observables are 
closely related. A correct theoretical description
of the three-nucleon system should be
able to reproduce both types of data.

The measurements were carried out using tensor-polarized deuteron
beams from the atomic beam polarized ion source~\cite{Cle95} at
the Triangle Universities Nuclear Laboratory (TUNL). The beams were
accelerated to $E_d=1.3$~MeV using the FN tandem accelerator, and
then directed into a 62-cm diameter scattering chamber.
Thin hydrogenated carbon targets were used~\cite{Black}
which consisted of approximately
$1\times 10^{18}$ and $2 \times 10^{18}$ hydrogen and carbon
atoms/cm$^2$, respectively. The deuteron beam loses $\approx 5$~keV
in these targets, leading to an average energy
of $E_{c.m.}=432\pm 3$~keV, where the error includes the uncertainty in
the incident energy.
The use of thin targets is very important at low energies for
minimizing energy loss and straggling effects.

The beam polarization was determined to $\pm 3$\% using the
${}^3{\rm He}$(d,p) reaction in an online polarimeter located behind the
scattering chamber. The polarimeter is described in Ref.~\cite{Ton80};
the calibration has been extended down to $E_d=1.3$~MeV.
This reaction is excellent for deuteron tensor polarimetry at low
energies due to the large cross section and tensor analyzing powers.
The data were taken using three spin states with tensor polarizations
$p_{ZZ}\approx\pm 0.7$ and $p_{ZZ}\approx 0$.
The spin states were cycled approximately once every second, in order
to minimize the effects of slow changes in beam position, target
thickness, or amplifier gain.
Tests for false asymmetries were carried out by measuring
${}^{197}$Au(d,d) scattering at $\theta_{lab}=40^\circ$ under identical
conditions as the p-d measurements.
For $E_d=1.3$~MeV, all of analyzing powers for ${}^{197}$Au(d,d)
are expected to be $<10^{-4}$~\cite{Kam85}.
The results were consistent with zero at the level
of $5\times 10^{-4}$, the statistical uncertainty of the measurement.

Scattered deuterons and protons were detected with silicon surface barrier
detectors located between 15 and 25~cm from the target. At certain angles
2- or 5-$\mu$m mylar foils were placed in front of the detectors to
stop heavy nuclei recoiling from the target,
or to separate the deuterons from
p(d,d) scattering from protons due to the ${}^{12}$C(d,p$_1$) reaction.
Backgrounds were typically less than 3\%, and were subtracted
using linear or exponential fits to background regions on either side of 
the peak of interest. The final errors include the uncertainty
in the background subtraction arising from the background region fitted,
the functional form used form the background, and the
analyzing power of the background.

Angular distributions of $T_{20}$ and $T_{21}$ are shown
in Fig.~\ref{fig:tens}. The error bars include contributions from
counting statistics, background subtraction, and dead time corrections,
but not the uncertainties in absolute beam polarization.

The theoretical methods developed in Ref.~\cite{KVR94} can
be applied to n-d as well as p-d scattering below the breakup
threshold. When realistic NN potentials and TNI
terms are considered, the corresponding results
allow for meaningful detailed comparisons with experimental data. 
Scattering waves up to orbital angular momentum $L=3$ have been taken 
into consideration in the construction of the scattering wave function.
The calculated $T_{20}$ and $T_{21}$ are shown in
Fig.~\ref{fig:tens}, and are seen to be in remarkably good agreement
with the experimental measurements.
A particularly stringent test of the theoretical approach is
provided by the $T_{20}$ data, whose complicated structure
is due to interference effects between the nuclear and Coulomb interactions.
The magnitude of the polarization observables gives a measure of the
importance of $L>0$ partial waves, for if only $S$-wave scattering
were considered, all of the polarization observables would be zero.

The calculations have been extended to lower energies in order to
make predictions for the effective range functions and N-d
scattering lengths.
In the past the doublet scattering length in N-d scattering has been 
calculated for a variety of interaction models~\cite{scatl}. 
A correct approach to the problem requires that the interaction 
model considered reproduces the three-nucleon binding energy. 
Apart from small differences between the various models,
a reasonable description is obtained
for the n-d system, in particular when the AV18+UR interaction is adopted. 
The calculated value $^2a_{nd}=0.63$~fm~\cite{KVR95}  
compares well with the experimental
value of $0.65\pm 0.04$~fm~\cite{ndexp}.
For the p-d case, all the calculations agree with the value
$^2a_{pd} \approx 0$.
Extraction of the experimental value is complicated by
the large curvature of the effective range function
when the energy approaches zero.

An effort is currently underway at TUNL to measure the p-d
differential cross section at $E_{c.m.}\approx 170$~keV
and $E_{c.m.}\approx 210$~keV~\cite{Black}.
For $E_{c.m.}<300$~keV the cross section is dominated by Coulomb
scattering and it is justified 
to consider the nuclear phase shift in only the $S$- and $P$-waves. 
The ${\cal R}$ matrices for
$J^\pi={1 \over 2}^+$ and ${3 \over 2}^+$ are then scalars.
Following Ref.~\cite{FRIAR89},
an effective range expansion for the corresponding element
$^JR_{00}={\rm tan}\delta^J_0$ can be performed. To be explicit, 
let us concentrate on the $S$-wave $J^\pi={1\over 2}^+$ phase shift.
For p-d scattering the effective range function is
\begin{equation}
 K(E)=C^2_0(\eta)k{\rm cot}\delta_0(k)+2k\eta h(\eta),
\end{equation}
where $\eta=2Me^2/3\hbar^2k$ is the Coulomb parameter,
$M$ the nucleon mass,
$C^2_0=2\pi\eta/({\rm e}^{2\pi\eta}-1)$,
$h(\eta)=-ln(\eta)+{\rm Re}\psi(1+i\eta)$,
and $\psi$ is the digamma function. 
Again, the calculations have been done using the AV18+UR nuclear
potential model and the Coulomb interaction.
Other electromagnetic terms of the AV18 potential
(such as the vacuum polarization and magnetic moment terms)
have been neglected in order to simplify the calculations,
since the asymptotic solutions can then be expressed in terms of the Coulomb 
functions. The inclusion of those neglected terms produces only small changes 
in the scattering lengths~\cite{AIP}.

The phase shift $\delta_0^{1/2}$
has been calculated for several energies with $E_{c.m.}<450$~keV.
The calculated values of the doublet effective range function
are plotted in Fig.~\ref{fig:range}, 
where the arrows indicate the energies at which the experiments were 
performed.
These numerical results reveal a pole in the effective range function
near zero energy. A good fit to the numerical results can be
obtained assuming
\begin{equation}
 K(E)= {-1/\,{}^2a_{pd}+\beta E \over 1+E/E_0}\ .
  \label{eq:ke}
\end{equation}
This singular behavior was anticipated in Ref.~\cite{FRIAR89}
where $S$-wave potentials were used
and it is corroborated here for realistic nuclear interactions.
The fitted values obtained for the free parameters are 
a doublet scattering length of ${}^2a_{pd}=0.024$~fm,
$\beta =-0.076$~fm${}^{-1}$~keV${}^{-1}$
and $E_0=3.13$~keV corresponding to the solid curve in Fig.~\ref{fig:range}.
A direct calculation for this quantity at zero energy 
gives ${}^2a_{pd}=0.027$~fm, in good agreement with the extrapolated result.
When all the electromagnetic terms of the AV18 potential are properly 
taken into account the value ${}^2a_{pd}=-0.022$~fm is obtained~\cite{KVR95}.
Since experiments for the p-d system below $E_{c.m.}=150$~keV are extremely
difficult, the form given in Eq.~(\ref{eq:ke})
is a useful guide for an energy-dependent
PSA intended to determine the p-d scattering lengths. 

The problems with extrapolation to $E=0$ are not present in the
$J^\pi={3\over 2}^+$ $S$-wave state.
The corresponding effective range function has a smooth
behavior near the N-d threshold~\cite{FRIAR89}.
The quartet scattering lengths, calculated using the AV18+UR model,
are $^4a_{nd}$=6.33~fm and $^4a_{pd}$=13.8~fm. 
Here again, for the n-d system the theoretical estimate
is quite close to the experimental value
$^4a_{nd}=6.35\pm 0.02$ fm~\cite{ndexp}.

Further information on the structure of the three-nucleon system 
can be obtained through the study of the asymptotic constants.
It is well known that the $J^\pi={1\over 2}^+$ elastic scattering
amplitude and the properties of the three-nucleon bound states are
closely related.
It is therefore of interest to compare the predictions of the AV18+UR
potential model to recent experimental results~\cite{Beata,Zeid,Knutson}.
For the absolute values of
the $S$-state asymptotic constant $C_S$, the $D$-state
asymptotic constant $C_D$ and the ratio $\eta=C_D/C_S$ we obtain:
     $C_S=1.854$, $C_D=0.0798$, $\eta=0.0430$ for $^3$H, and 
     $C_S=1.878$, $C_D=0.0752$, $\eta=0.0400$ for $^3$He.
From the above results it can be seen that the 
ratio $\eta$ is lower than previous calculations~\cite{Kamimura}
and agree well with the experimental results:
$\eta({}^3{\rm H })=0.0411\pm 0.0013\pm 0.0012$~\cite{Beata} and
                   $0.0431\pm 0.0025$~\cite{Knutson},
$\eta({}^3{\rm He})=0.0386\pm 0.0045 \pm0.0012$~\cite{Zeid}.
Since $\eta$ gives a measure of the
$D$-state amplitude, these results are expected,
to some extent, due to a weaker
tensor component present in the AV18 NN potential.

The study of the three-nucleon properties near the N-d 
threshold was the aim of the present paper.
The theoretical calculations agree well in magnitude and shape
with the highly-accurate $T_{20}$
and $T_{21}$ data at 432~keV where Coulomb effects are of
considerable importance. These results increase our confidence in the
theoretical p-d scattering wavefunctions used recently to
calculate the $S$-factor and analyzing powers
of the ${}^2{\rm H}(p,\gamma)$ reaction
at very low energies~\cite{cattura}.

The calculated $J^\pi={1\over 2}^+$ $S$-wave effective range function
for p-d scattering reveals a pole a few keV above the N-d threshold 
and predicts a doublet scattering length ${}^2a_{pd} = 0.024$~fm.
Both values are consistent with recent Faddeev calculations.
The calculated effective range function can serve as a guide
for future experiments intended to determine
the $a_{pd}$ scattering lengths.

The calculated $S$- and $D$-state asymptotic normalization constants
for ${}^3$He and ${}^3$H are in excellent agreement with recent measurements.  

\section*{Acknowledgements}

The authors would like to thank B.~J.~Crowe, W.~H.~Geist, and K.~D.~Veal
for their assistance in the data collection process.
One of the authors (A. K.) would like to thank Duke University and 
TUNL for hospitality and partial support
during his stay in Durham, where part of the present work was performed.
This work was supported in part by the U.S. Department of 
Energy, Office of High Energy and Nuclear Physics, under 
grant No. DE-FG05-88ER40442. 




\begin{figure}
\caption{The tensor analyzing powers $T_{20}$ and $T_{21}$
for p(d,d) scattering at $E_{c.m.}=432$~keV, as
a function of center-of-mass angle. The experimental data are shown as
circles and the solid curves show the theoretical calculations.}
\label{fig:tens}
\end{figure}

\begin{figure}
\caption{Effective range function values (solid circles) calculated
with the p-d doublet phase shifts at various energies.
The solid curve is the fit obtained
with Eq.~(\protect\ref{eq:ke}).
The arrows indicate the energies at which
experiments have been performed at TUNL.}
\label{fig:range}
\end{figure}

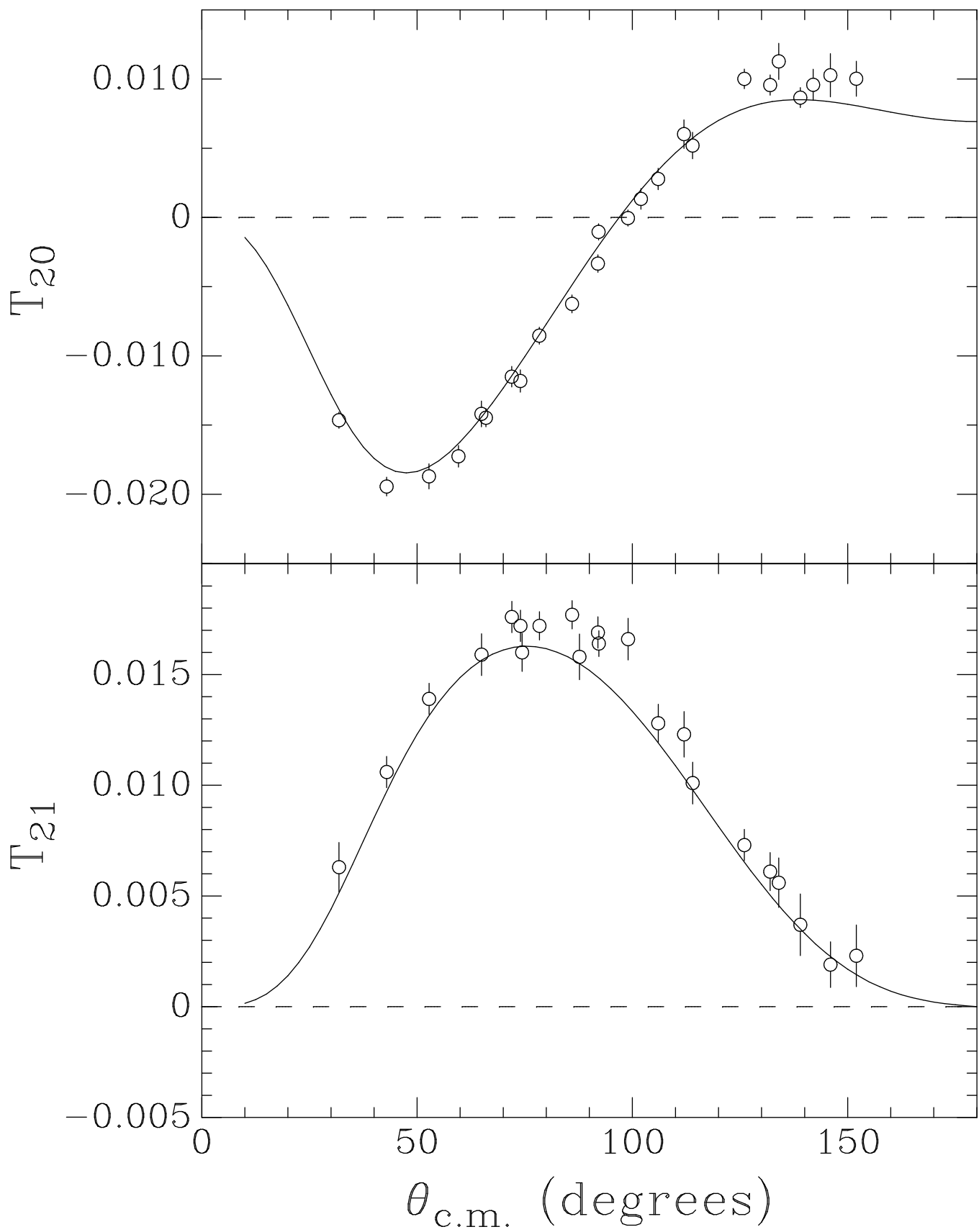
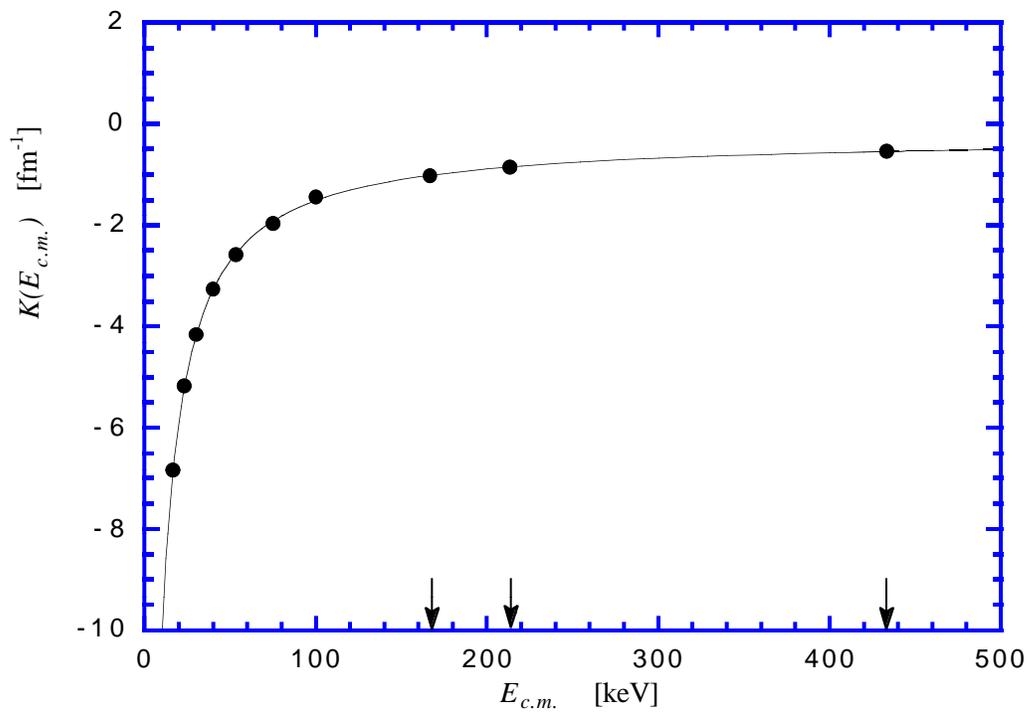
\end{document}